# The Discovery and History of the Dalgaranga Meteorite Crater, Western Australia

Duane W. Hamacher[1] and Craig O'Neill[2]

[1]Nura Gili Centre for Indigenous Programs, University of New South Wales, Sydney, NSW, 2052, Australia

Email: d.hamacher@unsw.edu.au

[2]Department of Earth & Planetary Sciences, Macquarie University, Sydney, NSW, 2109, Australia

Email: craig.oneill@mq.edu.au

## Abstract

The Dalgaranga meteorite crater, 100 km northeast of Yalgoo, Western Australia, was one of the first impact structures identified in Australia, the smallest isolated crater found in Australia, and the only confirmed crater in the world associated with a mesosiderite projectile. 17 years passed before the Dalgaranga meteorites were described in the scientific literature and nearly 40 years passed before a survey of the structure was published. The reasons for the time-gap were never explained and a number of factual errors about the discovery and early history remain uncorrected in the scientific literature. Using historical and archival documents, and discussions with people involved in Dalgaranga research, the reasons for this time gap are explained by a series of minor misidentifications and coincidences. The age of the crater has yet to be determined, but using published data, we estimate the projectile mass to be 500-1000 kg.

## Introduction

Discovered in 1921, the Dalgaranga meteorite crater (27° 38' 05" S, 117° 17' 20" E, Fig. 1), 100 km northeast of Yalgoo, Western Australia, was the one of the first impact structures recognised in Australia. At 24 m wide and 4.5 m deep, it is the smallest isolated impact crater in Australia, and the only confirmed terrestrial meteorite crater formed by a mesosiderite projectile.

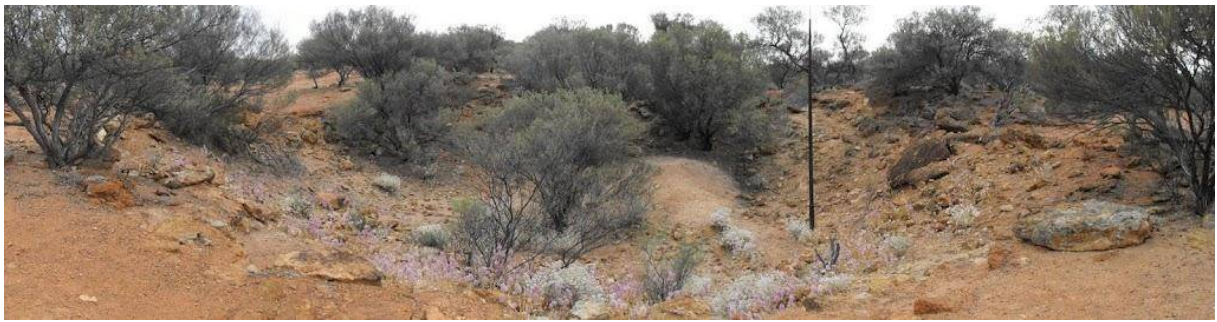

*Fig. 1: Panoramic image of Dalgaranga crater. From Megan Argo, taken November 2009.*

The age of Dalgaranga crater is not known, but estimates rang from 270,000 years using $^{10}$Be/$^{26}$Al exposure (Shoemaker *et al.* 1990) to < 3,000 years based on its well-preserved condition (Shoemaker & Shoemaker 1988). The lower limit would make it the youngest crater in Australia and among the ten youngest craters in the world (Grieve 1991).

Very little information about the discovery of the crater or the reasons why nearly 40 years passed before it was properly surveyed is reported in the scientific literature. This paper reveals connections between people and events that have gone unnoticed by the scientific community for nearly 90 years and corrects many factual errors that are commonly cited in the literature.

This paper uses historical documents collected from published literature, newspapers, unpublished archival materials, and personal communication with people involved in research at Dalgaranga. Documents include autobiographies and memoirs, biographies, contemporary newspaper articles using the Trove database, and unpublished letters, manuscripts, and other materials from the archives of the City Library of Carlsbad, the Western Australian Museum, the State Library of Western Australia, and the Library archives at Arizona State University. Family trees from Ancestry.com were used to help identify personnel involved in Dalgaranga research in cases where only initials were provided in the literature. Photos of important figures in the history of Dalgaranga crater research that have passed away are included where applicable.

**Discovery**

In the early 1920s, the 260,000-acre Dalgaranga station was owned by Alexander Robert Richardson (1847-1931; Fig. 2a) and managed by his nephew, Gerard Eardley Pierce Wellard (Fig. 2b). In 1983, Wellard published a book of memoirs that included his experiences at Dalgaranga (Wellard 1983:95-97). This book provides the following account of the crater's discovery and explains why so much time passed before the meteorites were reported in the literature.

An Aboriginal stockman named Billy Seward identified Dalgaranga crater in 1921. Seward (Fig. 2c) was mustering cattle on horseback when he and the horse nearly fell into the structure while in full gallop. He told Wellard and Wellard's brother about the "big hole" that he stumbled upon. Two days later, Seward took Wellard to the site, which according to Wellard, he immediately recognised as a probable impact crater. Wellard scanned the area for evidence of meteorite fragments and collected about 50-60 specimens, enough to "fill a gallon can" (Wellard 1983:95). Most of them were collected in the immediate vicinity of the crater, but some were found up to 300 meters away. He described a majority of the fragments as being a few centimeters wide. Wellard took the fragments back to his home on the station and there they sat for nearly two years.

It was not until 1923 that Wellard made an effort to have the specimens examined scientifically, although no reason is given for this. That July, Richardson visited his nephew at Dalgaranga station (Anon 1923). Wellard told his uncle about the crater and asked him to take the meteorites to the Western Australian Museum in Perth to be studied. Richardson agreed and took them back to his farm, Lowlands, on the Serpentine River near Perth (Anon 1931). He placed the bag of specimens in his office but soon forgot about them. For six months they sat unnoticed, until one day the 75 year-old Richardson stubbed his toe on the bag, which "jogged his memory enough" to remember that they needed to go to the museum but had forgotten they were from Dalgaranga Station (Wellard 1983:96).

Richardson took the bag of meteorites to the Department of Mines in Perth, which shared the same building as the Western Australian Museum. At the time Richardson dropped of the specimens, the Mines department was in the process of relocating to a new facility. During the moving process, according to Wellard, the bag of meteorites was misplaced. Eager to learn more about the meteorites his uncle took in for study, Wellard claims that he sent a number of inquiries about the collection to the museum - but never received an answer. In May 1924, Wellard travelled to Perth so his wife could give birth to their daughter, Rosemary (Anon 1924; Wellard 1983:163). While in Perth, he visited the museum and enquired about the meteorites. To his frustration, no one seemed to know anything about the stones or where they were located. Wellard tells us that he was eventually able to find a man from the Department of Mines who remembered receiving the bag of stones but did not know what they were or where they were from. The man reportedly said that the bag had been misplaced during the relocation and its whereabouts were unknown. This was apparently the last correspondence between Wellard and the museum regarding the meteorites. It is uncertain if Wellard's specimens were ever located (A. Bevan pers. comm. 2012).

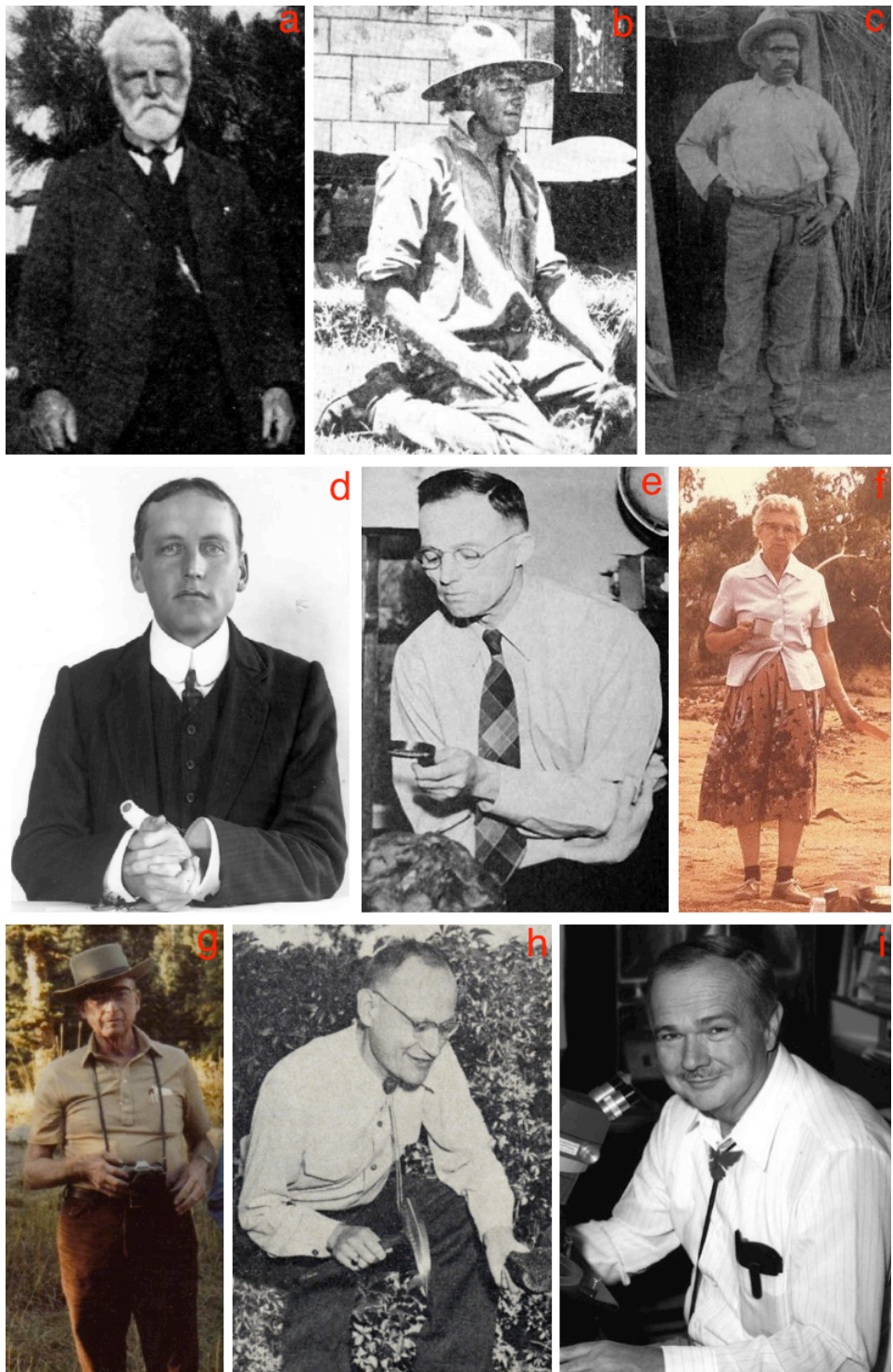

*Fig. 2: (a) Alexander Richardson (c.f. Wellard 1983:20), (b) Gerard Wellard circa 1925 (c.f. Wellard 1983:35), (c) Billy Seward circa 1915 (c.f. Wellard 1983:36), (d) Edward Simpson (Geological Society of Western Australia), (e) Harvey Nininger (c.f. Nininger 1972), (f) Addie Nininger circa 1959 (Carlsbad City Library), (g) Glenn Huss (Doris Banks), (h) Richard Pearl (c.f. Wilson 2012), (i) Eugene Shoemaker (Wikimedia Commons).*

**The Simpson Study**

The eventual study of meteorites from Dalgaranga was instigated by a chain of events, sparked by the occurrence of a rare phenomenon some six years after Wellard's visit to the museum. At 14:00 on 7 April 1930, a fireball streaked across the sky in the vicinity of Gundaring, WA (Simpson 1938). The fireball burst into two halves above the village of Cuballing, travelling in parallel paths and appearing to strike the Earth in the distance (*ibid*). The Government Astronomer, Harold Burnham Curlewis, used descriptions from observers to calculate the fireball's trajectory and estimate where it landed (Curlewis 1930). The incident generated substantial public interest and numerous articles about the fireball appeared in newspapers across Australia (Simpson 1938). Unfortunately, several searches of the fall-area estimated by Curlewis were unfruitful (Anon 1937).

In February 1937, W.G. Armstrong found a 53.5 kg IIC iron meteorite in the Kumarina copper field, approximately 430 km northeast of Dalgaranga (Grady 2000:285). Three months later, F. Quinn found a 112.5 kg IIIAB iron meteorite in Gundaring – within the fall-area predicted by Curlewis (Grady 2000:227). This was hailed in the press as the meteorite that formed the fireball in 1930[1] (Simpson 1938). The discovery of the Kumerina and Gundaring meteorites once again stirred up significant public interest in meteorites (Simpson 1938). Edward Sydney Simpson (Fig. 2d), the Government Mineralogist and Analyst for WA and one of the founders of the Western Australian School of Mines, displayed specimens of both finds at a meeting of the Royal Society of Western Australia in June 1937 (Anon, 1937). Simpson studied these and a dozen other meteorites from across WA, publishing his results in June 1938 (Simpson 1938).

One of the meteorites studied by Simpson was from Dalgaranga. According to Simpson (1938), Wellard contacted him about the meteorites he had found on his sheep station. Simpson explained that the original collection of specimens had been lost but Wellard was able to secure a 40 g fragment from the current station manager and gave it to Simpson (by this time Wellard no longer managed or resided at Dalgaranga). Simpson stated that Wellard's description of the crater confirmed rumors he had heard when he visited Yalgoo in 1932, although he had not visited the structure himself. In his memoirs, Wellard seemed pleased a meteorite from Dalgaranga had finally been studied and published. But he noted two inaccuracies reported in Simpson's paper. First, Simpson misspelled his name as "Willard", which would lead to a series of events many years later that were resolved by an incredible coincidence. Second, Simpson said that Wellard quoted the crater's diameter as "75 yards across" at the top, "50 yards at the bottom," and "15 feet deep". Wellard (1983:96) says the only measurement he made was by counting the number of steps he took to walk *around* the rim, which was about 225 feet. Therefore, it was actually ~70 ft (22 m) in width as opposed to ~70 m (230 ft). A diameter of 22 m is close to the accepted estimate of 24 m (Haines 2005:483). From Wellard's description that the northwestern side of the crater appeared to be thrust upward more than any other part of the rim, Simpson concluded that the impactor had travelled from the southeast. He classified the Dalgaranga meteorite as a medium octahedrite and hoped for a chance to visit the site to collect and analyze more specimens. Unfortunately, Simpson passed away from heart complications on 30 August 1939 before he was able to return to the site (Prider 1988).

---

[1] Years later, analysis of the Gundaring meteorite revealed that it had been exposed to the elements for quite some time, thus negating it as the remnant of the 1930 fireball (Buchwald 1975; McCall *et al.* 2006:310).

**The Nininger Expeditions**

For many years, Simpson's paper was the only original source of information regarding the crater and associated meteorites. At the time, no proper survey of the crater had ever been conducted or published. Therefore, the only record of the crater itself was the anecdotal account from Wellard reported by Simpson - and one of the important facts (the crater dimensions) was incorrect. Over the next several years, numerous researchers cited the Simpson paper without the crater or meteorites undergoing any further investigation (Nininger & Huss 1960:620).

Simpson's description of Dalgaranga eventually caught the attention of Harvey Harlow Nininger (Fig. 2e), an American entomologist and self-taught meteoriticist, commonly dubbed the Father of American Meteoritics (Palmer 1999). One subject of interest to Nininger was the explosive formation of meteorite craters. According to Nininger, no crater less than 28 m in diameter exhibited evidence of formation under explosive vaporization of the projectile. One diagnostic piece of evidence that could be used to test whether this occurred was the presence of metallic spheroids, which formed when the meteorite melted on impact and solidified as it cooled. Spheroids were found at Barringer (Meteor) crater in Arizona (D=1.2 km) and Odessa crater in Texas (177 m), but none were found at Haviland crater in Kansas (16.8 m) or the Sikhote-Alin craters in eastern Siberia (< 25 m). Many questions about the formation of these spheroids remained unanswered and Nininger believed a crater the size of Dalgaranga as reported by Simpson (70 m) could help shed light on the issue (Nininger & Huss 1960; Huss 1962).

A trip to Australia was necessary, but by the late 1950s Nininger had reached a state of financial crisis. In 1958, he sold 20% of his extensive meteorite collection to the British Museum for $140,000 U.S. – over $1.1 million U.S. in 2013 currency (Nininger 1972:204-206). This provided him with the funds to lead a 72-day expedition to Australia, New Zealand, Fiji, and Hawai'i with his wife, Addie (Fig. 2f), and amateur geologist Allan O. Kelly to study craters and search for meteorites and tektites. The Australian leg of the expedition took place over 17 days. The team flew to Perth in February 1959 and drove to Dalgaranga to survey the impact structure (Kelly 1961:145-148). This would be the first scientific investigation of the Dalgaranga meteorite crater – nearly 38 years after its discovery. With directions from a local shop-keeper named John L.S. "Jack" Nevill and the assistance of the current station manager, Cyril H. Ross (who purchased the lease in 1942; Anon 1998:6), his wife, and an Aboriginal employee, Nininger's team reached the crater on 9 February and set up camp near the crater rim (Kelly 1961:148; Nininger & Huss 1960:620-621). Upon first investigating the structure, they realised that the measurements reported by Simpson were in error (Nininger 1959a). To compound their frustration, the metal detector they brought would not function properly and they were reduced to looking for meteorites visually and searching for spheroids using magnets (Nininger & Huss 1960). During the survey, Nininger's team noticed a striking lack of meteorites in the area. They were only able to collect 23 specimens with a total mass of 149 g (e.g. Fig. 3). The largest single fragment weighed only 30 g. This stands in contrast to an account given by Peter Lancaster Brown, who claimed that "metallic fragments were not difficult to find" when he visited the crater in the early 1950s (Brown 1975:191). This suggests that prospectors may have removed most of the larger fragments as word of the crater spread.

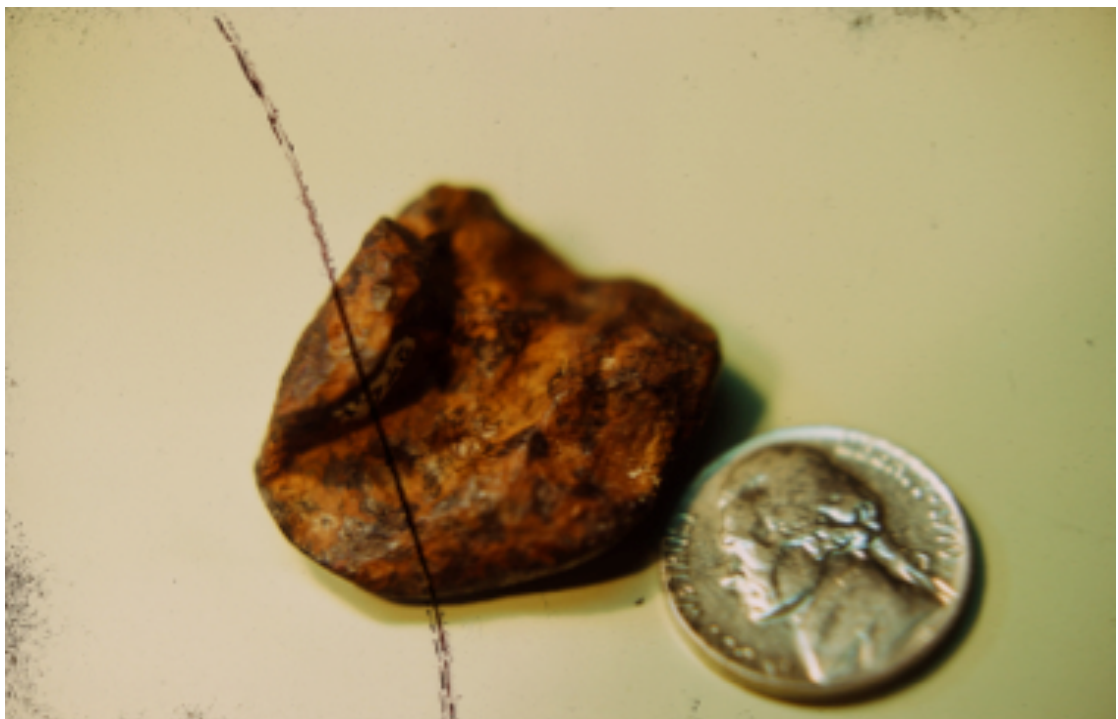

*Fig. 3: The meteorite fragment from Dalgaranga bent by the impact. Collected by Nininger and Kelly, pictured with a U.S. 5-cent coin for scale (Alan O. Kelly Collection, Carlsbad City Library, Photo 00010196).*

The crater appeared to be an explosion impact, caused not by vaporization but by violent fragmentation of the projectile (Nininger & Huss 1960). Since Simpson reported that the meteorite was nickel-iron in composition, the low number of fragments collected seemed to contradict the fragmentation hypothesis. The team re-considered that the projectile was stony-iron in composition and concluded that the main mass of the impactor must be buried within the crater (Nininger & Huss 1960:622). Attempts to find this mass failed. They decided to focus their efforts, given the short amount of time they had left, on collecting the material for future analysis. The team mapped the crater and the distribution of meteorites before leaving early the next afternoon (Fig. 4). They never published their maps and none appeared in the literature until 2005, compiled from a survey conducted in 1986 (Shoemaker *et al.* 2005). The team's maps are included in Kelly's unpublished book manuscript in the Carlsbad City Library, California (Kelly 1961:152). Kelly provides two maps of the crater: a cross section and an aerial view, which are published here for the first time (Fig. 5).

During the survey, Kelly found numerous Aboriginal flint flakes around the crater, which he thought might have marked places where meteorites had been recovered by Aboriginal people (Kelly 1961:153-154). Kelly noted that flint is not native to the area, so the flakes must have been brought in by human agency. He estimated that the impact would have occurred within the last few hundred years based on the degree of weathering and erosion on the crater walls and in the ejecta. In his mind, there was "little doubt" that Aboriginal people witnessed the fall, since it would have been visible from some distance, even in daylight, had it occurred during human habitation of Australia. He speculated that Aboriginal people had taken the meteorite fragments and left behind the flint flakes as "*an exchange gift for the fiery god that came out of the sky*" (Kelly 1961:153). No Aboriginal oral traditions of the Dalgaranga crater or meteorites are reported in the literature, but Bevan & Griffin (1994) believe that a meteorite fragment reportedly recovered near Murchison Downs, WA in 1925 was transported by human agency from Dalgaranga crater, some 200 km away. They suggest it may have been transported by Aboriginal people, but nothing more is known about how or why it was found so far from the impact site. It should be noted that the label on the Murchison Downs fragment gave the donor's name as "Richardson", although it is uncertain if this is a reference to Wellard's uncle, A.R. Richardson (A. Bevan pers. comm. 2012).

On 13 February, the team came upon an Aboriginal camp near Leonora, WA , where many of the locals had collected tektites, which they colloquially called "meteorites" (Kelly 1961:162). Nearly everyone at the camp knew about them and Nininger was able to purchase two samples. Apparently, Aboriginal people had been collecting them for years to sell to museums or white collectors. Kelly explains that as the demand for tektites waned many of the Aboriginal people simply lost or discarded their collections (Nininger 1972:213).

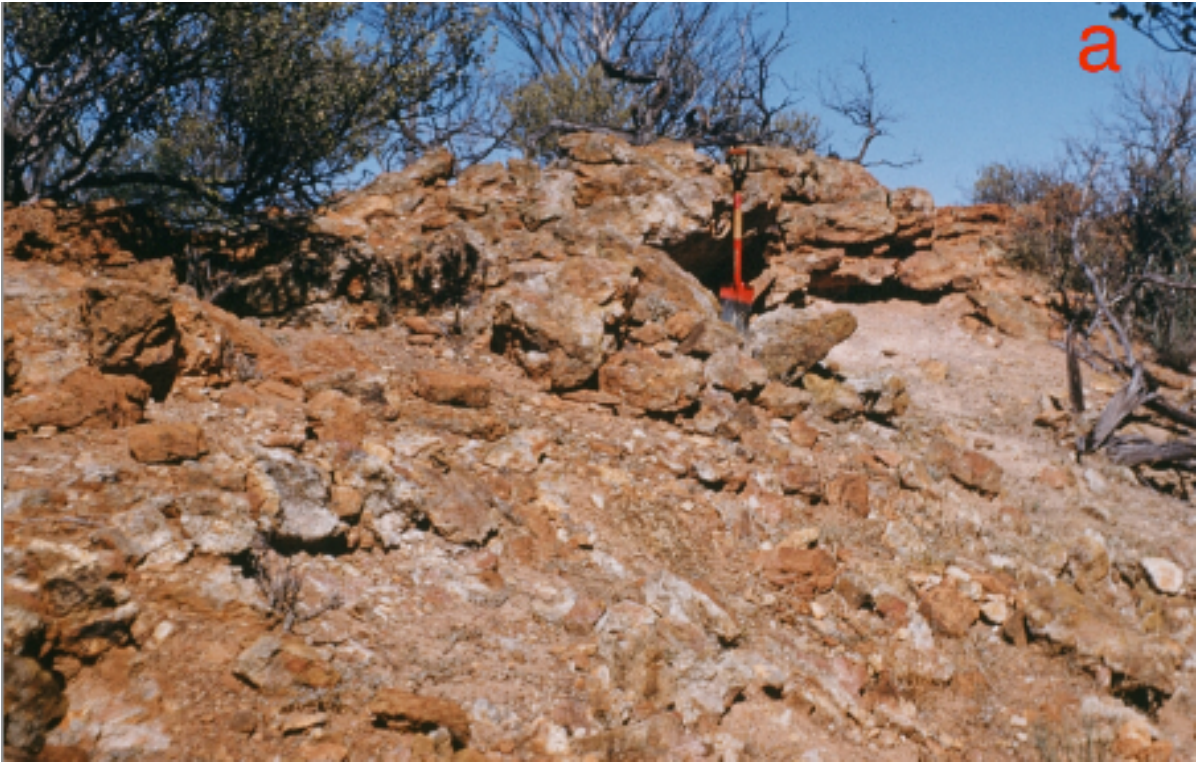

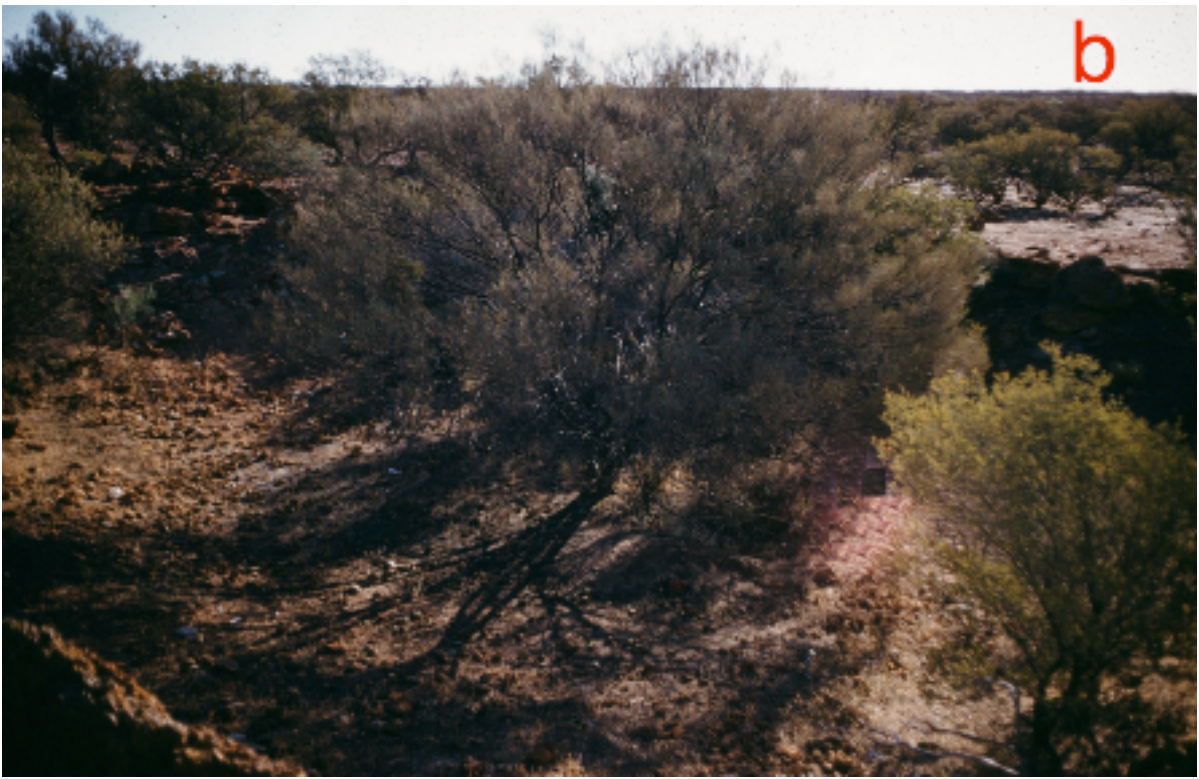

*Fig. 4: Previously unpublished images of Dalgaranga crater from the 1959 Nininger & Kelly Expedition. (a) The rim of the crater as seen from the focus with a shovel for scale, (b) A large mulga bush growing in the crater's focus before it was cut down in 1959 (Alan O. Kelly Collection, Carlsbad City Library, photos 00010235 and 00010124, respectively).*

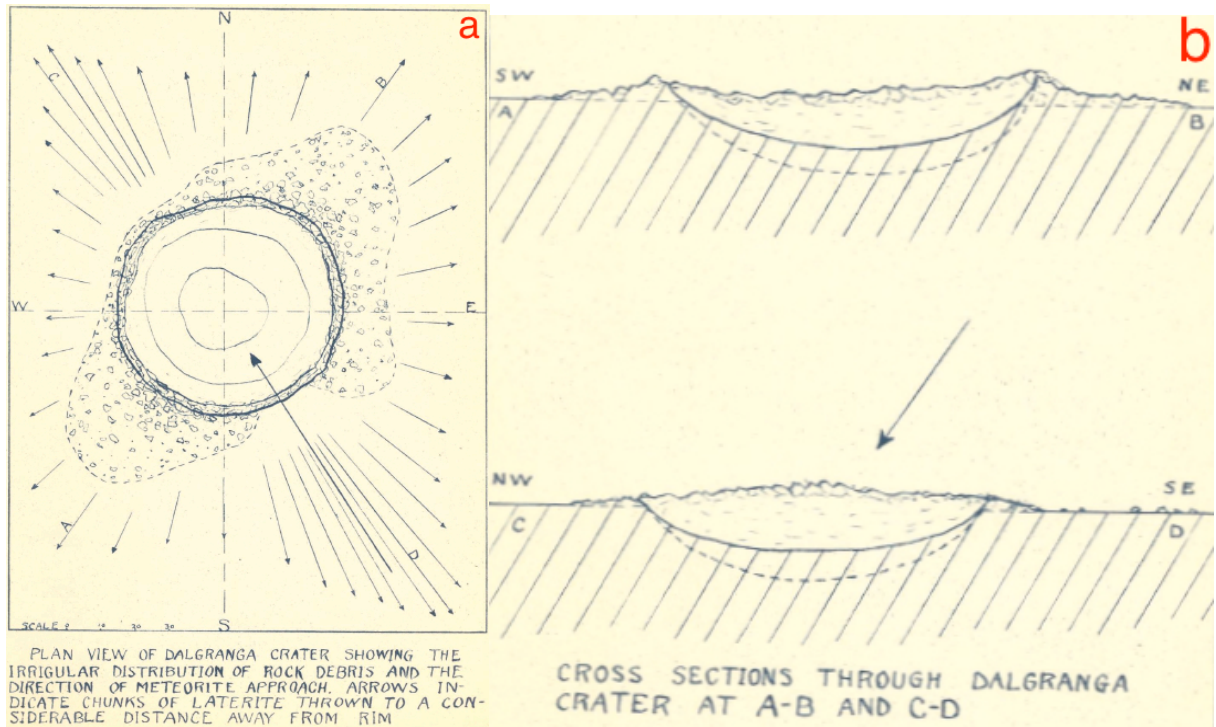

*Fig. 5: Maps of Dalgaranga crater produced by Nininger and Kelly in 1959 (c.f. Kelly 1961:166-167), courtesy of the Carlsbad City Library. (a) A bird's eye view of the crater and ejecta blanket. The arrow indicates the proposed trajectory of the impactor. (b) Cross-section of crater and breccia fill across points C and D. The arrow indicates the proposed impact angle of the meteorite (Top: southwest to northeast. Bottom: northwest to southeast).*

Because the expedition covered such a large area, several months passed before Nininger was able to properly study the specimens from Dalgaranga. When he did, he realised these were not octahedrite (Ni-Fe) meteorites as classified by Simpson, but rare mesosiderite (stony-iron) meteorites (Nininger & Huss 1960). This was significant as no other impact crater in the world was associated with this type of meteorite. Upon discovering this, Nininger realised that they had relied too heavily on Simpson's paper (*ibid*:623). The combination of their tight schedule, the intense heat, and the various technical issues meant that they had missed an opportunity to investigate the site in detail. This would have aided in answering pertinent questions about the impact cratering process, so Nininger was eager to return to Dalgaranga to conduct a more in-depth survey.

Nininger organised the second expedition to Australia and on 18 October 1959, Nininger arrived at Dalgaranga with his son-in-law, Glenn I. Huss (Fig. 2g) a geologist and director of the American Meteorite Laboratory in Denver, Colorado (Nininger 1972:213). They remained at the site for nearly two weeks, making detailed measurements and maps of the crater, collecting 207 additional specimens, clearing the trees within the crater, and excavating the crater floor to search for the main projectile they thought lay within (Nininger & Huss 1960:627-628; Huss, 1962a,b; McCall & de Laeter, 1965:28; Cleverly, 1962). The results of both surveys were published in 1960 (Nininger & Huss 1960). Huss (1962b) described the crater as square in shape, similar to Barringer "Meteor" crater in Arizona and described the symmetry of the crater in some detail.

According to Huss (1962b), he and Nininger noted that the crater displayed a bilateral symmetry. At an azimuth of 289° (northwest), the rim lay flat and revealed almost no rim rubble. The rim tilted increasingly until at points 90° on either side, the rim was 40° to the horizontal before leveling out again to almost flat at a point just south of due east. The highest concentration of ejecta corresponded with the areas where the rim displayed the greatest tilt. The ground some 15-20 m from the southwest rim showed a fairly heavy layer of coarse, disintegrated granite. The northeast rim revealed a large quantity of large blocks of laterite. A thick layer of ejected laterite extended 45-60 m to the north-northeast and east side of the crater (*ibid*). Based on their geological mapping, it seems the impactor struck the edge of a small mound or hillock of laterite.

Among their findings was the realization that the projectile consisted of mainly stone (90%) with nests of mesosideritic (stony-iron) and sideritic (iron) materials. Nininger also concluded that the impactor, which he and Huss estimate had a mass of 10-20 tonnes, had not undergone vaporization but instead impacted at a relatively low speed, completely shattering the meteorite and creating the shallow crater (Nininger 1972:215; Nininger & Huss 1960:639). Nininger and Huss claimed that the reason so few meteorite fragments remained is that much of the stony components had eroded away, leaving behind iron components.

Huss cut down the largest mulga bush in the centre of the crater and counted the growth rings, which numbered 54 (Huss 1962b:14). This means the tree germinated around 1905 - roughly consistent with Simpson's estimate based on the bush's height of 4.5 m as cited by Wellard. The tree had germinated on top of a 1.6 m layer of breccia fill and no fill had accumulated around the roots, showing conclusively that the age of the crater far exceeded the age of the tree. Examining the rocks on the crater wall and the fill exposed from the excavation of the crater floor, Nininger and Huss estimated that the crater formed some 25,000 years ago.

There was no communication between Wellard and Nininger's team during the expedition. Wellard had long since sold the lease for Dalgaranga station and moved 700 km away to Yladgee station near Gnowangerup, about 300 km southeast of Perth (Kelly 1961:149; Barunha 1934). In fact, Nininger & Huss (1960) and Huss (1962b) cite Wellard's name as "Willard" and include the same inaccuracies that were recorded in the Simpson paper, simply because they did not know this information was in error.

Nearly 40 years after the crater's discovery, it was finally surveyed and the true nature of the Dalgaranga impactor was reported in the scientific literature. Further studies of the meteorites and crater would be undertaken (e.g. McCall 1977; Consolmagno & Britt 1995; Smith & Hodge 1996; Hidaka & Yoneda 2011; Shoemaker *et al.* 1990, 2005). But a year after Nininger & Huss' paper was published, an extraordinary coincidence occurred that linked everyone involved.

**An Extraordinary Coincidence**

In August 1961, Wellard and his wife, Katherine (neé Clifton), decided to take a trip to England on the Royal Dutch Mail Line ship *Johan van Oldenbarnevelt* (Wellard 1983:97). After three days at sea, Wellard's wife became engaged in a conversation with a "Mrs. Pearl". The woman mentioned that her husband was a professor of geology in the United States with a special interest in meteorites. Mr. Wellard, who had joined the conversation, mentioned that he also had an interest in meteorites. The woman was sure her husband would like to meet him. A few days later, Wellard met up with the woman's husband, a "Professor Pearl". Pearl explained that he and his wife had flown to Perth from America for the sole purpose of studying a special meteorite and visiting a unique impact crater called Dalgaranga to meet the man who found the meteorite, a "Mr. Willard". He conceded that while had studied the meteorite, he was very disappointed that he was unable to locate Willard. Somewhat taken aback, Wellard told Pearl that he was now talking to the very person he was unable to find! Pearl promised to send Wellard a copy of the report on the Dalgaranga meteorite he had studied, which he honored to the joy of Wellard.

The identity of "Professor Pearl" was not given in Wellard's memoirs. Further investigation reveals that this was in fact Professor Richard Maxwell Pearl (Fig. 2h) from Colorado College in Colorado Springs (Wilson 2012). Professor Pearl was well known for his studies of minerals, gems, and meteorites and was a member, and later Fellow, of the Society of Research on Meteorites, which later became the Meteoritical Society (Bostick 2002). Richard and his wife, Mignon, were friends and colleagues of Harvey Nininger. Pearl and Nininger had collaborated on various projects and it is now evident how Pearl came to learn of Dalgaranga. Glenn Huss published a two-part account of the second survey for which he participated (Huss 1962a,b). He confirmed that in early 1962 Wellard had met Richard and Mignon Pearl on the ocean liner to Amsterdam.

Information from Wellard's book is included on a tourist sign at the crater, although many pieces of the story we have provided here are not included and some of the details on the sign are inaccurate. It is noteworthy that Dalgaranga was not Wellard's final encounter with newly discovered meteorites. Around 1976, a 33.6 kg meteorite was discovered near Gnowangerup, WA by a road worker and given to Wellard. He donated it to the Western Australian Museum in October 1979 (de Laeter 1982:137).

After Nininger and Huss published their survey of Dalgaranga crater, a "Mr. Latham" conducted a follow-up excavation on the southern side of the crater floor. Latham sent two small specimens to Jack Nevill, who sent the fragments to F. William G. Power a mine inspector and WWII POW from Geraldton who forwarded them to geochemist William H. Cleverly, also at the School of Mines and an honorary associate at the Western Australian Museum (Bevan 1997). It is apparent from correspondence that Jack Nevill obtained some 20 kg of meteorites from Dalgaranga, which he shipped to Nininger in Sedona, Arizona in 1959 (Nininger 1959b). Samples of the material collected by Latham and Nevill are currently located in the Western Australian Museum (catalogue number WAM 12365) and the Western Australia School of Mines in Kalgoorlie, WA.

**The McCall Expedition**

In September 1963, University of Western Australia geologist Gerard J. H. "Joe" McCall led a two-day expedition to Dalgaranga, accompanied by Edward P. Henderson, curator at the U.S. National Museum, and E. Car, a biologist at the Western Australian Museum (McCall 1963). McCall notes that Nininger's measurements were accurate, citing a rim-to-rim diameter of 21.3 m, and a maximum 3.2 m deep from the rim to the top of the breccia fill. McCall (1963) noted that the granite and laterite separated by a layer of "thin shaly iron" showed a regular outward pit of ~30° from the crater centre. He then described the sediment within the crater, which had not previously been published. Using the pit dug by Nininger, McCall explains that the crater is infilled by a two-fold unconsolidated deposit composed of soil and granite and laterite boulders at the top and a meter of course stratified grits of granite detritus at the base (possibly of Aeolian origin). McCall did not reach the bottom of these sediments using the second pit dug by Nininger, which had a depth of 1.37 m. Based on the depth and nature of the sediment, McCall reports that the crater is of considerable age, though he does not provide an estimate. He cites the slope of the wall of granite rock at 50°-60° and notes that the rim is higher to the East with no marginal raised mound above the level of the surrounding plain to the North. McCall's favorite experience (McCall pers. comm. 2012) was the 1969 Australian Institute for Mining & Metallurgy (AusIMM) meeting in Cue, WA. As part of the meeting, a session was held at the crater in which the attendees sat along the rim while McCall spoke from the crater's focus.

The primary goal of the expedition was to collect meteorite fragments, but pickings were slim and the McCall team was only able to collect a handful of "small chips of iron." Henderson was allowed to take all of the recovered material on the condition that a specimen of mesosiderite material be given to McCall for analysis. In 1965, McCall published the first thin-section analysis of a Dalgaranga mesosiderite, using specimens collected during the survey (McCall 1965). Given the mineralogical composition of the fragment, McCall estimated that the projectile was only 1-2 tonnes (for reasonable higher impact velocities, and assuming no atmospheric fragmentation), not 10-20 tonnes as quoted by Nininger & Huss (1960).

William Cleverly and Maitland Keith Quartermaine visited the crater in 1962 (Cleverly 1962). According to Cleverly, Simpson's claim that the impactor came from the southeast was unsupported by the apparent excavations conducted by Nininger, since they seemed to be in the opposite direction in the crater. Nothing about this trip or survey was published in the literature except a photo of the crater, which was published in Bevan (1996:426).

**Site Protection**

On 25 July 1962, Cleverly (1962) wrote a letter to William D.L. "David" Ride, then Director of Western Australian Museum and President of the Royal Society of Western Australia, encouraging him to lodge an application to establish the crater and the surrounding area as a protected reserve. In the letter, Cleverly pointed out several areas of concern regarding the preservation of the crater and of the material within. These included fire hazards by the mulga scrub, damage caused by prospectors and researchers, and litter from visitors. Much of this is compounded by the fact that the crater is very small and that further excavations could erase important evidence regarding the history of the fill-material.

The application was successful and the crater was declared a Ministerial Temporary Reserve (3232H) through the Western Australia Department of Mines on 9 June 1965 (Cleverly 1962; A. Bevan pers. comm. 2012). Dalgaranga is registered as a protected site under the Mines Department and permission to collect must to be gained from the Minister for Mines. Meteoritic material is registered in the collection at the Western Australian Museum in accordance with the Museum Act 1969. The Dalgaranga crater is now classified as a State Geoheritage Reserve (Grey *et al.*, 2010; GSWA Record 2010/13).

**The Shoemaker Study**

In 1986, Eugene Shoemaker (Fig. 2i) and his astronomer-wife, Carolyn, visited Dalgaranga crater and conducted an in-depth study of the structure and associated meteorites (Shoemaker *et al.* 1990; 2005). The Shoemakers mapped the site and excavated the interior, using the original pit dug by Nininger near the eastern edge of the crater floor. The bilateral symmetry of the ejecta is consistent with laboratory experiments of low angle impacts (Gault & Wedekind 1978), suggesting the projectile impacted at a low incidence angle of 10-15° from the horizontal (Shoemaker *et al.* 2005:542).

The age of the Dalgaranga impact remains a conundrum, with estimates varying by two orders of magnitude. Simpson (1938) noted that it must have occurred prior to 1910, based on the height of the mulga bush growing in the middle of the crater. Kelly (1961) suggested a young age of just a few hundred years. Nininger & Huss (1960:639) cite an age of less than 50,000 years, nearer to 25,000 years, although they did not use any quantitative techniques. Nininger & Huss collected charcoal from the crater rim and within the excavated pit for dating but never published the results. McCall (1963) claimed the structure was of "considerable age" based on sedimentation within the crater. Shoemaker *et al.* (1990) conducted a $^{10}Be/^{26}Al$ exposure age test on samples collected from the top of the granite bedrock near the crater rim, giving an age of 270,000 years. This fit a low model erosion rate comparable to other parts of the Australian landscape. However, breccia block samples taken from the lower part of the crater walls have relatively high $^{10}Be$ and $^{26}Al$ abundances. This indicates the samples were exposed on the surface prior to the impact, thus complicating the use of this technique. Shoemaker *et al.* proposed to use the silicate phase of a sample meteorite for $^{14}C$ testing, which they believed would give a more accurate estimate. The results were never published. Upon surveying the crater in the late 1980s, the Shoemaker's believed the preservation of subtle morphology in the ejecta suggested an age of less than 3,000 years (Shoemaker *et al.* 2005:542). Despite these efforts, the age of the Dalgaranga impact is still unknown.

**Size and Mass Estimates**

The mass of the Dalgaranga impactor is also not well known, with early estimates varying by an order of magnitude. Estimates of 10,000-20,000 kg (D=1.7-2.1 m) by Nininger & Huss (1960) were based on a lack of specimens expected for a fragmentation crater of this size, leading them to believe is was mostly stone in composition. McCall (1965) noted the poor recovery of meteoritic material from the site, and on comparison with other fragmentation craters, suggested a mass of around 1,000-2,000 kg, or 0.7-1.0 m in diameter. We must remember that Brown (1975) claimed that meteorite fragments were not hard to find in the early 1950s, indicating that prospectors may be partially responsible for the lack of recoverable material by the time McCall reached the site in the 1960s. In 1996, Toby R. Smith, Paul W. Hodge, Alex W. R. Bevan, and Jenny Bevan surveyed the crater. Smith & Hodge (1996) analyzed soil samples from Dalgaranga to study the small meteoritic particles that are formed as the projectile passes through the atmosphere and strikes the Earth. The high abundance of these particles in the soil at Dalgaranga, their size (0.4-0.1 mm), and the fact that they are heavily weathered and unmelted is consistent with fragmentation as opposed to vaporization. Based on the mass of the fragments, Smith & Hodge estimate the impactor mass was on the order of 4,000 kg – double the upper estimate of McCall (1965).

An impactor of the order of 1 m in diameter is likely to have lost over half of its kinetic energy during its passage through the atmosphere, making final estimates of its size uncertain (Melosh 1989). However, we can apply a Pi-Group scaling approach for sensible estimates on the impactor and target rock to obtain some quantitative assessment of the robustness of the previous estimates. Using this approach, the final diameter of the impactor $L$ is derived using Collins *et al.* (2005, their equations 21 and 22), given by:

$$L \approx \left[ \frac{D_f}{1.451 \cdot \left(\frac{\rho_i}{\rho_t}\right)^{1/3} \cdot v_i^{0.44} \cdot g_E^{-0.22} \cdot \sin^{1/3} \theta} \right]^{1.282} \quad \text{(Eqn. 1)}$$

Here $D_f$ is the final crater diameter (24 m), $\rho_i$ is the impactor density, assumed to be 4,250 kg m$^{-3}$ for a mesosiderite (Britt & Consolmagno 2003), $\rho_t$ is the target rock density (2,650 kg m$^{-3}$ for Archaean granite of the Pilbara craton; McCall 1965), $v_i$ is the impactor velocity (4,000-22,000 m s$^{-1}$; Melosh 1989), $g_E$ is Earth's gravitational acceleration, and $\theta$ is the angle of incidence (assumed to be low angle for Dalgaranga based on crater morphology – we explore a range from 5° to 20°).

The effect of these parameters is shown in Fig. 6 for varying impact velocity and impact angle. Assuming this scaling is valid for a small fragmentation crater like Dalgaranga, estimates for the impactor's size range from 0.4-1.1 m in diameter. The mass of the impactor, if a mesosideritic density is assumed, is between 200-3,000 kg at the extreme, with the most probable bounds between 500-1,000 kg, roughly consistent with the estimates given by McCall (1965) and Smith & Hodge (1996).

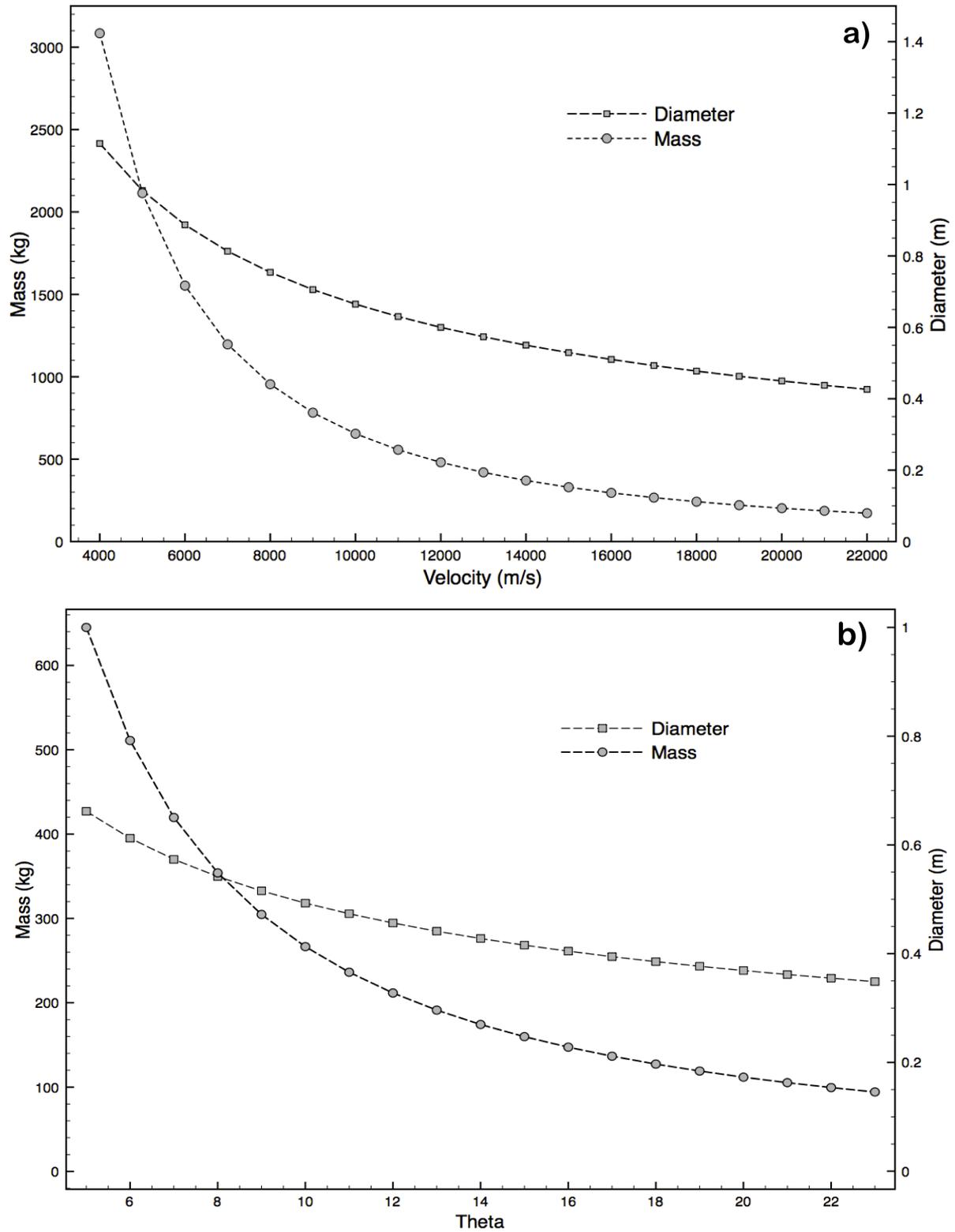

*Fig. 6. (a) Effect of impact velocity on the final estimate of mass (left scale) and diameter (right scale), calculated using Equation 1 for an assumed impact angle of 10°, and all other parameters as discussed in the text. (b) Effect of varying impact angle on final estimates of the diameter and mass of the impactor, assuming an impact velocity of 17 km s$^{-1}$.*

**Summary**

An Aboriginal stockman, named Billy Seward, discovered Dalgaranga crater in 1921 and meteorite fragments were recovered by the station manager, Gerard Wellard. Wellard sent specimens to Perth to be investigated but they were lost. Seventeen years passed before a proper study of Dalgaranga meteorite was published, but some of the information about the crater itself was in error. A further nineteen years passed before a survey of the crater was published and we have pieced together the reasons for this. We provide early maps of the crater and numerous historical details about its study that have not been previously published. We explain how early mistranslations in data about the crater led to many errors in estimates about its size. To date, the mineralogy of the Dalgaranga impactor is well understood but estimates of its mass have varied widely. Using pi-scaling equations, we estimate it is on the order of 500-1,000 kg. Estimates of the crater's age vary by three orders of magnitude, from a few hundred years to a few hundred thousand years. Its age remains a mystery and we encourage future work to help solve this piece of the puzzle.

**Acknowledgements**

We thank Amy Davis, Gary Huss, Peggy Schaller, Alex Bevan, Joe McCall, and John Goldsmith for their assistance, advice, and comments. We are indebted to the Carlsbad City Library (California), Arizona State University Library, the State Library of Western Australia, the Geological Society of Western Australia, and the Western Australian Museum for archival materials. This research made use of the National Library of Australia's Trove archival database (trove.nla.gov.au), the NASA Astrophysics Data System (adsabs.harvard.edu), and Ancestry.com. We would also like to acknowledge the referees for their helpful and useful comments. Additional photos from the Nininger & Kelly expedition will be published as a supplemental paper in AJES, courtesy of the Carlsbad City Library.